# Loss and Coupling Tuning via Heterogeneous Integration of MoS₂ Layers in Silicon Photonics

*R. Maiti[1], C. Patil[1], R. Hemnani[1], M. Miscuglio[1], R. Amin[1], Z. Ma[1], R. Chaudhary[2,3], A. T. C. Johnson[2,3], L. Bartels[4], R. Agarwal[2], V. J. Sorger[1] \**

[1]Department of Electrical and Computer Engineering, George Washington University, Washington, DC 20052, USA

[2]Department of Materials Science and Engineering, University of Pennsylvania, Philadelphia, PA 19104, USA

[3]Department of Physics and Astronomy, University of Pennsylvania, Philadelphia, PA 19104, USA

[4]Chemistry and Materials Science and Engineering, University of California, Riverside, California 92521, USA

*Corresponding Author E-mail: sorger@gwu.edu*

## Abstract

Layered two-dimensional (2D) materials provide a wide range of unique properties as compared to their bulk counterpart, making them ideal for heterogeneous integration for on-chip interconnects. Hence, a detailed understanding of the loss and index change on Si integrated platform is a prerequisite for advances in opto-electronic devices impacting optical communication technology, signal processing, and possibly photonic-based computing. Here, we present an experimental guide to characterize transition metal dichalcogenides (TMDs), once monolithically integrated into the Silicon photonic platform at 1.55 μm wavelength. We describe the passive tunable coupling effect of the resonator in terms of loss induced as a function of 2D material layer coverage length and thickness. Further, we demonstrate a TMD-ring based hybrid platform as a refractive index sensor where resonance shift has been mapped out as a function of flakes thickness which correlates well with our simulated data. These experimental findings on passive TMD-Si hybrid platform open up a new dimension by controlling the effective change in loss and index, which may lead to the potential application of 2D material based active on chip photonics.

**INTRODUCTION:**

Atomically thin two-dimensional (2D) materials research has progressed rapidly after the isolation of graphene in 2004 [1-4]. While graphene shows many exceptional properties, its lack of an electronic bandgap has stimulated the search for another graphene-like 2D material [5-8]. Transition metal dichalcogenides (TMDs), a family of materials with a general formula of $MX_2$, where M is a transition metal (Mo, W, Re) and X is a chalcogen (S, Se or Te), provide a promising alternative to integrate them into atomically precise heterostructures combining conducting (graphene) and insulating (hBN) 2D materials. A stable member of the TMDs family, molybdenum disulfide ($MoS_2$), has attracted widespread attention for a variety of next-generation electrical and opto-electronic properties such as high room temperature mobility, high switching characteristics, bandgap tunability, and high exciton binding energies [11-13]. Bulk $MoS_2$ has an indirect bandgap of ~1.2 eV which, due to quantum confinement, crosses over to a direct bandgap of ~1.8 eV, when the material is a monolayer [12]. Due to this tunable electrical and optical properties, $MoS_2$ could be a prospect for future advances in the field of nano-optics and photonics [13,14]. This material is studied here as monolithically integrated with Silicon photonics as just one example, where we are interested in the respective impact of optical absorption and index-shift whence heterogeneously added to a Silicon photonic waveguide and microring resonator (MRR). Beyond $MoS_2$, other TMDs could be studied as well in follow-up work on the same or similar integrated photonics platform, which is interesting, since the spectral distance of each TMD's exciton relative to the waveguide probing wavelength (here, $\lambda = 1550$ nm) is different; thus, the real vs. imaginary part index decay away from the exciton resonance has a different impact on the telecom-operating photonic structures.

Silicon photonics is becoming an integration platform of large interest for optical datacom and telecom applications [15]. However, Silicon's weak electro-optic properties and indirect bandgap severely limit opto-electronic device functionality. In contrast, hybrid or heterogeneous photonic integration solutions offer an appealing approach, when combined with an optical low-loss, yet commercially accessible large volume and low-cost CMOS fabrication technology such as Si/SiN photonics [16-19]. Other active opto-electronic materials such as transparent conductive oxides, while showing high switching performance, usually introduce relatively high optical losses [20]. Whereas, because of the advent of sufficiently strong van der Waals (vdW) force, 2D materials can (in principle) be easily integrated with photonic chip, offering a rich variety of electronic and optical properties that enable light generation, modulation, and detection could be a promising platform for next-generation PIC [21-24]. In reality, the state-of-the-art of TMDs transfer techniques is not benign with taped-out chip technology due to the inability to place a single 2D material flake on the pre-fabricated photonics chip without incurring significant cross-contamination (e.g. transfer of undesired flakes). We recently provided a solution for this challenge developing a 2D material printer enabling cross-contamination-free transfers without impacting the underlying photonic waveguide structures reported in ref [25].

Here, we demonstrate a novel heterogeneous platform to study the physical properties of TMDs by newly developed 2D printer transfer technique, enabling rapid and precise transfer of 2D atomic layers on the integrated photonic chip without any cross contamination. Using the TMD-Silicon heterogeneous integrated platform, we perform a comparative study to determine the optical loss and refractive index change as a function of 2D material Silicon waveguide and microring resonator (MRR) coverage length and TMD ($MoS_2$) thickness at a telecom wavelength. The effect of MRR-to-waveguide coupling has been mapped out in terms of resonance shift as a function of monolayer coverage analyzing the loss induced by monolayer $MoS_2$ which is found to be about 0.005 dB/μm. We obtain a resonance shift per unit waveguide coverage length of 0.064 nm/μm as a function of thickness which matches our numerical results well. Together these experimental studies of integrating $MoS_2$ with Silicon photonics shows an induced negligible loss, but relatively strong index-tuning potential thus paving the way for future studies of active opto-electronic device technology.

**METHODS:**

Here, we demonstrate a heterogeneous platform to study the effect of ultrathin TMDs towards building on-chip active device component. The study is performed on taped-out Si photonic chips (APPLIED NANOTOOLS INC.). It is important to keep all the physical parameters unchanged before and after the transfer of the 2D materials to single out the influence solely from atomic layered materials, thus; first, we coat a uniform layer of polymethyl methacrylate (PMMA) (~300 nm) as a cladding for the improvement of coupling efficiencies of the grating couplers (GC). Then, in order to keep a similar coupling efficiency and to eliminate the variation of light coupling, it is important to remove the PMMA layer and transfer the 2D material over the targeted opening areas of the devices. A box shape opening has been made by electron beam lithography for transferring 2D materials on top of the photonic devices (Fig. 2a).

In order to understand the potential of 2D TMDs at Telecom wavelength range, here we study $MoS_2$ integrated Si photonic platform as a function of layer thickness. Few-layer $MoS_2$ flakes are obtained using scotch tape exfoliation technique, whereas the monolayers are triangular flakes grown on $Si/SiO_2$ substrate by CVD process [26, 27]. First, we transfer those flakes onto an intermediate PDMS substrate via a KOH assisted wet chemical etching step. Then, the precise transfers of TMD materials are performed by using our developed 2D printer method very efficiently [2D Printer Video]. Briefly, this method comprises of a micro stamper to transfer the material from intermediate PDMS which is transparent and can be aligned under a microscope precisely at any desired device location via micro-positioners. We need to scan over a PDMS (Gel pack) substrate to find a flake of proper dimensions so that it could be transferred at a targeted location without having any cross contamination. Thereafter, the micro-stamper guides a flake to the target location and transfers it onto the substrate provided the effective contact area of the stamper is greater than the flakes area. After the transfer is complete, the devices are tested for their optical transmission and spectral shift again to determine the effect caused by the TMDs layer as a function of coverage length on the waveguide and MRR and as a function of TMD thickness. The experimental setup for measuring the hybrid TMD-Si devices consists of a tunable laser source (Agilent 8164B) and a broadband source (AEDFA-PA-30-B-FA) from where light is injected into the grating coupler optimized for the TM mode propagation in the waveguide. The light output from the MRR is coupled to the output fiber by a similar grating coupler and detected by a detector or an optical spectral analyzer (OSA202). The fundamental modes of the waveguide which integrate the $MoS_2$ layer of different thickness are found from the finite element method (FEM) simulations using the mode analysis tool in Comsol Multiphysics.

**RESULTS & DISCUSSION**:

In order to realize $MoS_2$ as an active material at telecom wavelength (here 1.55 µm), it is imperative to understand the interaction of monolayer and few layers system on integrated Si photonic devices (Figure 1). We study the loss and coupling in detail as a function of monolayer coverage and thickness of the $MoS_2$. It is well-known that the optical properties of TMDs materials are dominated by excitons: bound electron-hole pairs with strong binding energy due to quantum confinement and weak screening of the Coulomb interaction [12]. For any active integrated device structures, one needs to characterize the individual material system systematically such as loss or effective index change at the targeted device operation wavelength range. Here, we investigate the loss impact and impact on the effective mode index of the Si waveguide hundreds of nanometers away from the excitonic transition of $MoS_2$ (A exciton ~1.88 eV & B exciton ~2.06 eV) [13]. We anticipate a minimal loss impact but meaningful index change upon heterogeneous integration. While we are not modulating the TMD here electrically, the passive impact of the modes index and impact on an MRR provide fundamental insides in the potential of TMD-Si hybrid devices such as phase modulators [28-30]. We, therefore, study the waveguide bus-to-ring coupling change by shifting the MRR's phase upon adding $MoS_2$. We measure the corresponding transmission spectra which show a coupling change as compared to bare ring indicating MRR mode index tuning (Figures 1c&f). The improvement of coupling can be attributed as shifting the MRR from the over coupled regime towards the critically coupled regime by inducing loss which are evident since the quality factor decreases from ~1500 to ~1100 as a function of the flake thickness, suggesting gradual increase of loss mainly due to scattering effect, caused by small impedance mismatch between bare and TMD covered sections of the ring.

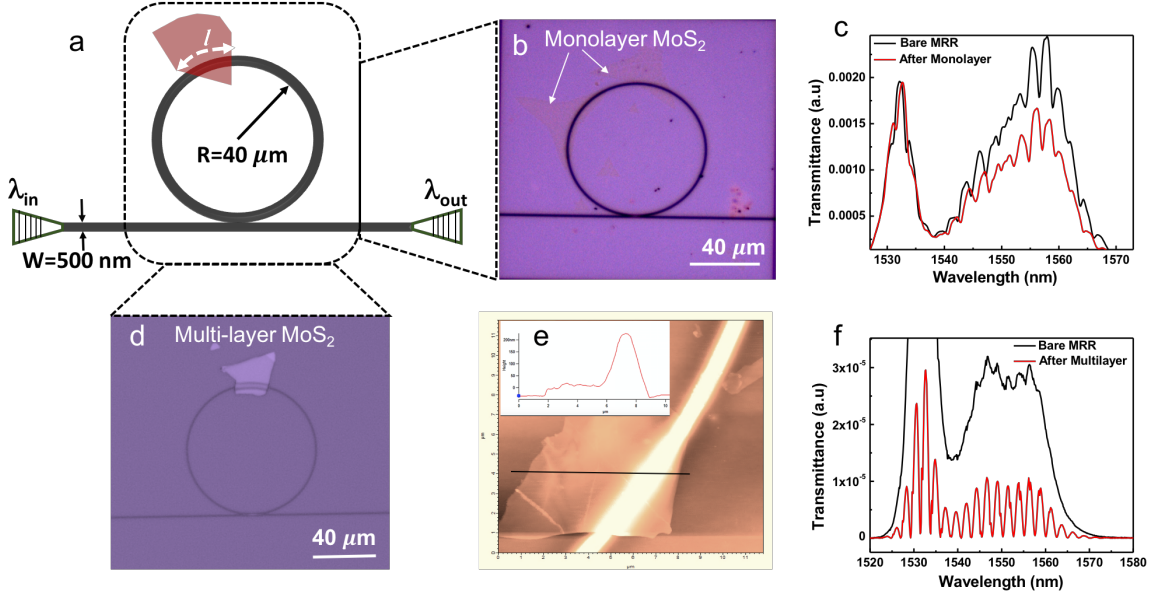

**Fig. 1. MoS₂ loaded micro-ring resonator (MRR). a**, Schematic b, optical microscope image of an MRR (R= 40 μm & W= 500 nm) covered by two monolayer MoS₂ flakes with coverage lengths ($l_1$ and $l_2$) precisely transferred using our developed 2D printer technique [25]. c, Transmission output before and after the transfer of monolayer MoS₂ showing improvement of coupling efficiency. d & e, optical microscope and AFM image of a MRR showing a multi-layer MoS₂ flake with coverage length 22 μm and thickness of 30 nm, respectively. f, Transmission output before and after the transfer of multilayer MoS₂ flakes which display a gradual increase of visibility suggesting the improvement of coupling efficiency as it brings the device close to critically coupled regime after the transfer of the TMDs layer.

Quantitative modeling and analyzing the MRR's resonance change, the fringe-visibility can be optimized (shifting towards critical coupling) in two ways: either by increasing coverage length or by increasing the thickness of flakes. So, to understand the coupling effect, it is important to extract coupling coefficients, especially round-trip transmission coefficients ($a$) as a function of coverage. The transmission, $T$, from an all-pass MRR (Figure 3a) is given by,

$$T_n = \frac{a^2 + r^2 - 2ar\,cos\varphi}{1 + r^2 a^2 - 2ar\,cos\varphi} \qquad (1)$$

where $\varphi$ is the round-trip phase shift, $r$ is the self-coupling coefficient and $a$ is round-trip transmission coefficient related to the power attenuation coefficients by,

$$a^2 = \exp\big(-\alpha_{Si}(2\pi R - l)\big) * \exp\,(-\alpha_{TMD-Si} * l) \qquad (2)$$

where $l$ = TMD coverage length, R is the radius of MRR, $\alpha_{Si}$ and $\alpha_{TMD-Si}$ are the linear propagation losses for Si waveguide and the TMD-transferred portion of the Si waveguide in the ring, respectively. To obtain the propagation loss more quantitatively we deploy the cutback method and integrate for mono-few layers of $MoS_2$ on linear waveguides of different lengths and measure the relative transmission (Figures 2b &c). The loss can be attributed as a combination of absorption and the scattering effect. Since, the popagation wavelength through the waveguide is well below the bandgap of $MoS_2$, the loss due to absorption is negligible. Hence, to minimize the effect of scattering and find the effect of absorption, we transfer single flakes with increasing coverage lengths (Figure 2e). The propagation losses obtained from the linear fit are is 0.008 dB/μm, 0.005 dB/µm and 0.04 dB/µm, respectively for Si, monolayer and multi-layer flakes, respectively (Table 1).

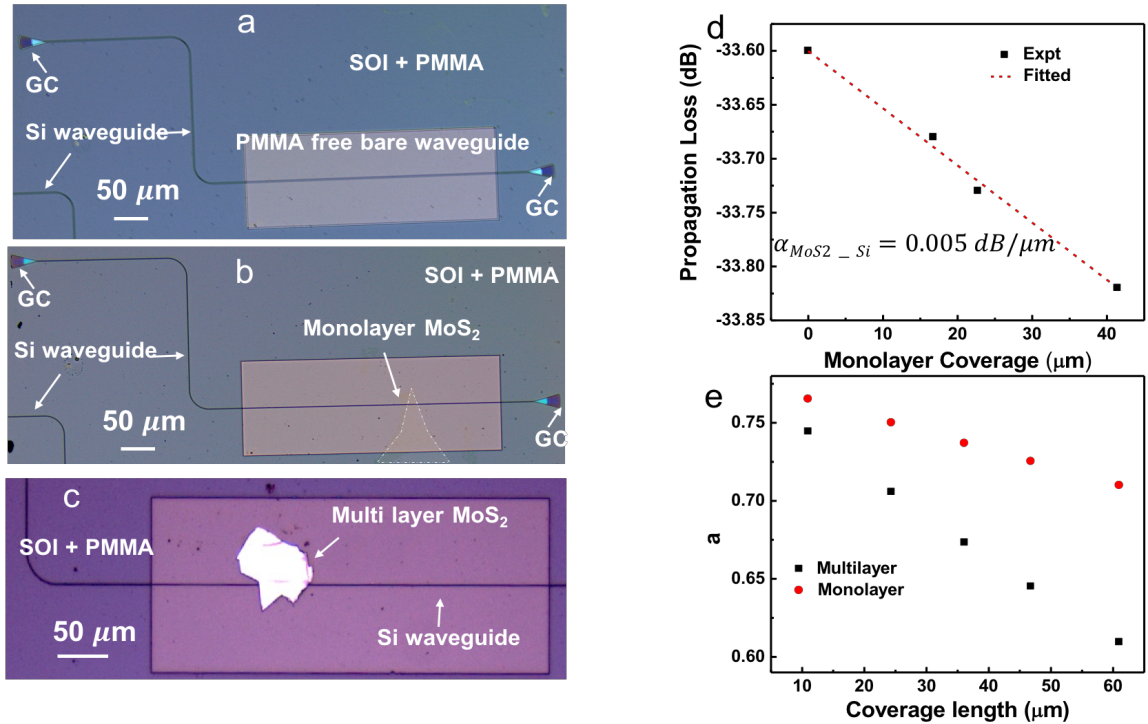

**Fig. 2. $MoS_2$ loaded Linear waveguides**. a, optical microscope image of the waveguide with an opening (400 μm x 100 μm) for TMD transfer to keep all physical parameters unchanged during the optical measurement. b & c, optical microscope image of waveguide covered by a monolayer CVD grown and a multilayer exfoliated flake with coverage length 10 μm and 50 μm, respectively. d, Optical loss output as a function of monolayer coverage length originates mainly due to the edge scattering effect. The propagation loss ($\alpha_{TMD-Si}$) for a TMD-covered portion of the ring is found to be 0.005 dB/μm using cutback measurement. e, Tunability of round-trip transmission coefficients explains the coupling improvement as a function of coverage length.

| Materials | Thickness | Band gap(eV) | Loss (dB/um) | Ref. |
|---|---|---|---|---|
| Graphene-Si waveguide | Monolayer | Zero | 0.1 | [30] |
| Black Phosphorus | 11 nm | ~1.5 | 0.2 | [22] |
| Black phosphorus | 100 nm | ~0.5 | 3.34 | [22] |
| $MoTe_2$-Si waveguide | Multi layer (~50 nm) | ~0.9 | 0.4 | [31] |
| $MoS_2$-Si waveguide | Multi-layer(~30-40 nm) | ~1.3 | 0.037 | This Work |
| $MoS_2$-Si waveguide | Monolayer | ~1.8 | 0.005 | This work |

**Table 1:** Table of Losses associated with 2D materials with heterogeneously integrated Si Platform

Inserting these values into (2), we find the round-trip transmission coefficients (*a*) to be tuned as a function of TMD coverage (Figure 2d). The variation of *a* from 0.77 to 0.60 as a function of coverage for multilayer flakes and from 0.75 to 0.70 for monolayer flakes, respectively. The result explains the improvement of visibility inducing the transition from over-coupled to towards critically-coupled regime since *a* = 1 is the zero-loss condition of the ring. Hence, the loss of tunability in MRRs can be manipulated accordingly by controlling the coverage length and thickness. We conclude that for a given device, one can determine the coverage length and thickness by configuring the device at a critically coupled condition for optimized light-matter-interaction [31].

Si-based MRRs provide a compact and ultra-sensitive platform to find sensitive detection of an unknown analyte for various applications [32, 33]. Here, the detection mechanism is mostly based on the change of the refractive index in the top-cladding of the MRR. This change can be sensed by the evanescent tail of the propagating optical mode governed by the effective mode index($n_{eff}$) and can be translated into resonance shift ($\Delta\lambda$). We observe the resonance shift of 1.5 nm upon increase of monolayer coverage length from 10 to 60 μm (Figure 3a) showing monotonic red shift which can be explained as follows: the resonant wavelength ($\lambda_{res}$) of a MRR is proportional to the effective refractive index of the propagating mode in the circular waveguide [33]. Therefore, the change in effective the mode index ($\Delta n_{eff}$) after transfer of MoS$_2$ flakes as compared to bare ring is related to a change in resonance shift ($\Delta\lambda$) by following eqn. $\Delta n_{eff} = \frac{\Delta\lambda}{\lambda_{res}} * n_{eff,bare}$, where, $n_{eff,bare}$ is the effective mode index for bare waveguide. The effective index for the bare waveguide can be found from FEM Eigenmode analysis choosing the TM-like mode in correspondence with our TM-grating designs used in measurements.

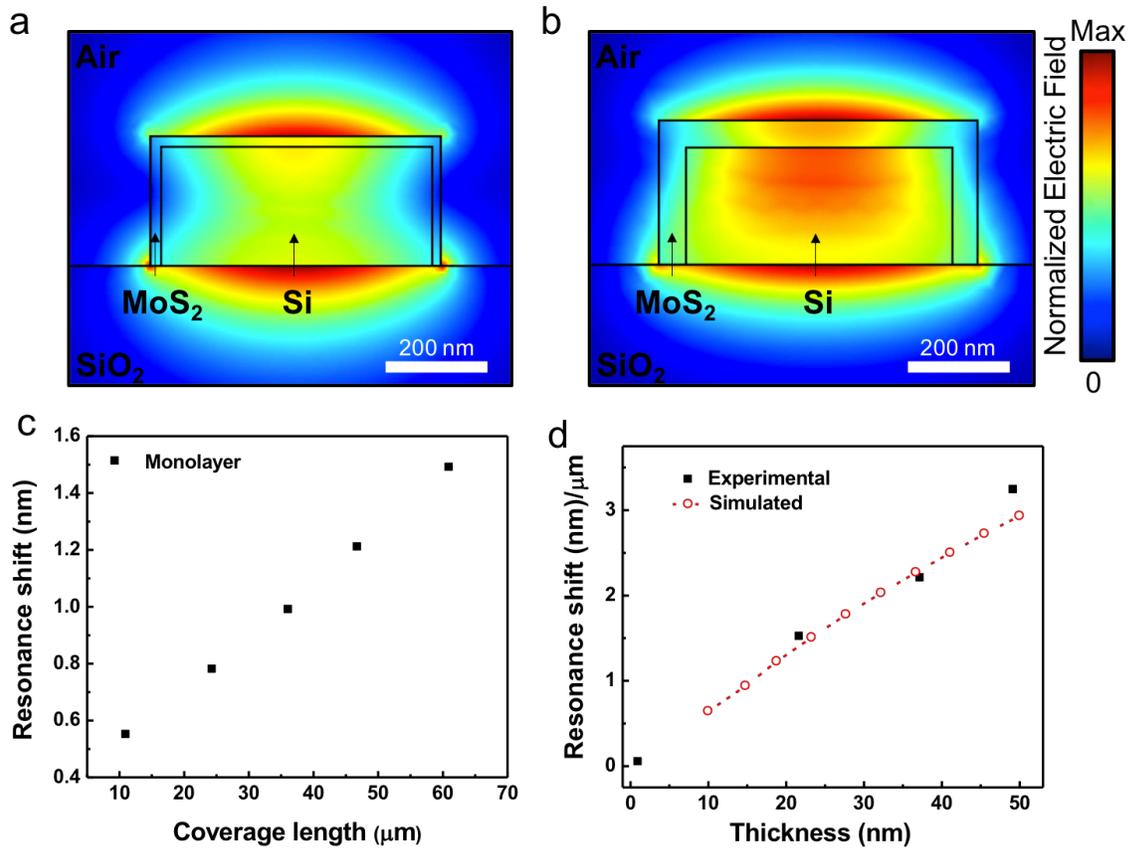

**Fig. 3. Ring resonator as refractive index sensor.** Mode profile (electric |E|-field), obtained through eigenmode analysis, for the portion of the ring with MoS$_2$ transferred flakes a, 20 nm and b, 50 nm thick (Scale bar 200 nm). Variation of Resonance shift ($\Delta\lambda$) and effective index change ($\Delta n_{eff}$) extracted from

c, as a function of MoS$_2$ coverage length and d, thickness, which is fully corroborated by a numerical simulation study.

Since, here the MRR is partially covered by MoS$_2$ flakes the effective refractive index of the ring can be formulated as an effective length-fraction index via [35],

$$n_{eff,ring} = \frac{(2\pi R - l)*n_{eff.bare} + l*n_{eff}}{2\pi R} \qquad (3)$$

where $R$ is the radius of the ring and $l$ is the MoS$_2$ coverage length. The effective index for monolayer covered waveguide is found to be 1.723 for TM mode. The effective index ($n_{eff}$) for flakes with different thickness can be found using FEM Eigenmode analysis. The refractive index of the MoS$_2$ layer was taken from [34] and Si refractive index from [35]. Figures 3a &b represent the normalized in-plane (|E$_x$|) electric field distribution for TM mode along the device which integrates a layer of MoS$_2$ of 20 and 50 nm thickness, respectively. It is possible to observe that the higher intensity of the electric field in correspondence of the MoS$_2$ layer and the consequently decreased leakages in air and substrate suggest a higher confinement for multi-layer MoS$_2$ (50 nm). At this stage, using eqn (3), we can obtain $\Delta n_{eff}$ and hence the resonance shift ($\Delta\lambda$), which is showing an unequivocal correlation with our experimental data (Figure 3d). We map out the resonance shift ($\Delta\lambda$) as a function of MoS$_2$ flake thickness (Figure 3d, (i)) upto 50 nm and observe a resonance shift per coverage length of 0.064 nm/μm, which is beyond the resolution limit of our spectrometer (~0.03 nm). The results indicate mono-multilayer MoS$_2$ integrated on Si photonic platform could be an interesting choice for active modulator devices at telecom wavelength.

**CONCLUSION:**

We have demonstrated the interaction between mono to multi-layers of MoS$_2$ heterogeneously integrated onto a Silicon photonic waveguides and microring cavities. The coupling regime of the ring can be tunable from over coupled regime to under coupled regime. The underlying physical mechanism of tunable coupling can be explained by extracting different coupling and loss coefficients as a function of coverage length and thickness. This study demonstrates a method to determine critical coverage for a given ring resonator which is an important parameter for obtaining maximum extinction ratio for the active modulator. We have mapped out resonance shift as a function of monolayer coverage and thickness of MoS$_2$ flakes which shows a resonance shift of 0.64 nm/μm correlates well with our simulated result. These findings along with the developed methodology for placing MRRs into critical coupling for active device functionality and determining the refractive index of 2D materials could be useful tools in future heterogeneous integrated photonic and opto-electronic devices.

## Acknowledgment

V.S. and L.B. are supported by AFOSR under award number FA9550-17-1-0377 and NSF Materials Genome Initiative under the award number NSF DMREF 14363300/1455050. R.A. and V.S. are supported by ARO under award number W911NF-16-2-0194. R.C. and A.T.C.J. acknowledge support through NSF EFRI 2-DARE award number 1542879.

**References**

[1] A. K. Geim and K. S. Novoselov, "The rise of graphene," Nat. Mater. 6(3), 183–191 (2007).

[2] S.V. Morozov, K.S. Novoselov, M.I. Katsnelson, F. Schedin, D.C. Elias, J.A. Jaszczak and A.K Geim, "Giant intrinsic carrier mobilities in graphene and its bilayer," Phys. Rev. Lett. 100 016602 (2008).

[3] M. Freitag, "Graphene: nanoelectronics goes flat out," Nat. Nanotechnol. 3 455–7 (2008).

[4] R. Amin, Z. Ma, R. Maiti, S. Khan, J.B. Khurgin, H. Dalir and V.J. Sorger, "Attojoule-efficient graphene optical modulators," App. Opt. 57, D130-140 (2018).

[5] S.Z. Butler, et al., "Progress, challenges, and opportunities in two-dimensional materials beyond graphene," ACS Nano 7, 2898–2926 (2013).

[6] M. Xu, T. Lian, M. Shi and H. Chen, "Graphene-like two-dimensional materials," Chem. Rev. 113, 3766–3798 (2013).

[7] Q. H. Wang, K. Kalantar-Zadeh, A. Kis, J. N. Coleman, and M. S. Strano, "Electronics and optoelectronics of two-dimensional transition metal dichalcogenides," Nat. Nanotechnol. 7(11), 699–712 (2012).

[8] R. Mas-Ballesté, C. Gómez-Navarro, J. Gómez-Herrero and F. Zamora, "2D materials: to graphene and beyond," Nanoscale 3, 20–30 (2011).

[9] S. Manzeli, D. Ovchinnikov, D. Pasquier, O.V. Yazyev and A. Kis, "2D transition metal dichalcogenides," Nat. Rev. Mat. 2, 17033 (2017).

[10] K. F. Mak, C. Lee, J. Hone, J. Shan, and T. F. Heinz, "Atomically Thin MoS2: A New Direct-Gap Semiconductor," Phys. Rev. Lett. 105(13), 136805 (2010).

[11] B. Radisavljevic, A. Radenovic, J. Brivio, V. Giacometti, and A. Kis, "Single-Layer MoS2 Transistors," Nat. Nanotechnol. 6(3), 147–150 (2011).

[12] M.M. Ugeda, A.J. Bradley, S.F. Shi, H. Felipe, Y. Zhang, D.Y. Qiu, W. Ruan, S.K. Mo, Z. Hussain, Z.X. Shen and F. Wang, "Giant bandgap renormalization and excitonic effects in a monolayer transition metal dichalcogenide semiconductor," Nat. mat. 13, 1091 (2014).

[13] A. Splendiani, L. Sun, Y. Zhang, T. Li, J. Kim, C.Y. Chim, G. Galli and F. Wang, "Emerging photoluminescence in monolayer MoS2," Nano letters. 10, 1271-1275 (2010).

[14] S. Mukherjee, R. Maiti, A. Midya, S. Das and S.K. Ray, "Tunable direct bandgap optical transitions in MoS2 nanocrystals for photonic devices," ACS Photon. 2, 760-768 (2015).

[15] D. A. Miller. Attojoule optoelectronics for low-energy information processing and communications. Journal of Lightwave Technology. 35, 346-396 (2017).

[16] S. Wu, S. Buckley, A.M. Jones, J.S. Ross, N. J. Ghimire, J. Yan, D.G. Mandrus, W. Yao, F. Hatami, J. Vučković, A. Majumdar and X. Xu, "Control of Two-Dimensional Excitonic Light Emission via Photonic Crystal" 2D Mater 1, 011001 (2014).

[17] X. Gan, Y. Gao, K. Fai Mak, X. Yao, R. Shiue, A. van der Zande, M. Trusheim, F. Hatami, T. Heinz, J. Hone and D. Englund, "Controlling the spontaneous emission rate of monolayer MoS2 in a photonic crystal nanocavity" Appl. Phys. Lett. 103, 181119 (2013).

[18] T.K. Fryett, K.L. Seyler, J. Zheng, C.-H. Liu, X. Xu, A. Majumdar, "Silicon Photonic Crystal Cavity Enhanced Second-Harmonic Generation from Monolayer Wse2," 2D Mater 4, 015031 (2017).

[19] G. Wei, T.K. Stanev, D. A, Czaplewski, I. W. Jung and N. P. Stern, "Silicon-nitride photonic circuits interfaced with monolayer MoS2," Appl. Phys. Lett. 107 (9), 091112 (2015).

[20] R. Amin, R. Maiti, C. Carfano, Z. Ma, M.H. Tahersima, Y. Lilach, D. Ratnayake, H. Dalir and V.J. Sorger, "0.52 V-mm ITO-based Mach-Zehnder Modulator in Silicon Photonics," arXiv:1809.03544 (2018).

[21] Y.Q. Bie, G. Grosso, M. Heuck, M.M. Furchi, Y. Cao, J.B. Zheng, D. Bunandar, E. Navarro-Moratalla, L. Zhou, D.K. Efetov, T. Taniguchi, K. Watanabe, J. Kong, D. Englund and P. Jarillo-Herrero, "A MoTe2-based light-emitting diode and photodetector for silicon photonic integrated circuits," Nat. Nanotechnol. 12, 1124 (2017).

[22] N. Youngblood, C. Chen, S. J. Koester and M. Li, "Waveguide-integrated black phosphorus photodetector with high responsivity and low dark current," Nat Photon (2015).

[23] Y. Ye, Z.J. Wong, X. Lu, X. Ni, H. Zhu, X. Chen, Y. Wang and X. Zhang, "Monolayer excitonic laser," Nat. Photon 9, 733 (2015).

[24] B. Lee, W. Liu, C.H. Naylor, J. Park, S.C. Malek, J.S. Berger, A.T.C. Johnson and R. Agarwal, "Electrical Tuning of Exciton–Plasmon Polariton Coupling in Monolayer MoS2 Integrated with Plasmonic Nanoantenna Lattice," Nano Lett. 17, 4541−4547 (2017).

[25] R.A. Hemnani, C. Carfano, J.P. Tischler, M.H. Tahersima, R. Maiti, L. Bartels, R. Agarwal and V.J. Sorger, "Towards a 2D Printer: A Deterministic Cross Contamination-free Transfer Method for Atomically Layered Materials," 2D Materials, 6, 015006, (2018).

[26] D. Kim, D. Z. Sun, W. H. Lu, Z. H. Cheng, Y. M. Zhu, D. Le, T. S. Rahman, L. Bartels, "Toward the growth of an aligned single-layer MoS2 film" Langmuir, 27, 11650, (2011).

[27] G. H. Han, N. J. Kybert, C. H. Naylor, B. S. Lee, J. Ping, J. H. Park, J. Kang, S. Y. Lee, Y. H. Lee, R. Agarwal & A. T. Charlie Johnson, "Seeded growth of highly crystalline molybdenum disulphide monolayers at controlled locations", Nat. Comm. 6 6128 (2015).

[28] C.T. Phare, Y.H. Lee, J. Cardenas, M. Lipson, "Graphene electro-optic modulator with 30 GHz bandwidth" Nat. Photon. 9, 511 (2015).

[29] I. Datta, S.H. Chae, G.R. Bhatt, B. Li, Y. Yu, L. Cao, J. Hone and M. Lipson, "Giant electro-refractive modulation of monolayer WS2 embedded in photonic structures," CLEO (2018).

[30] Y. Ding, X. Zhu, S. Xiao, H. Hu, L.H. Frandsen, N.A. Mortensen and K. Yvind, "Effective electro-optical modulation with high extinction ratio by a graphene–silicon microring resonator," Nano lett. 15, 4393-4400 (2015).

[31] R. Maiti, R. Hemnani, R. Amin, Z. Ma, M. Tahersima, T.A. Empante, H. Dalir, R. Agarwal, L. Bartels, V.J. Sorger, "Microring Resonators Coupling Tunability by Heterogeneous 2D Material Integration," arXiv:1807.03945, (2018).

[32] I.M. White and X. Fan, "On the performance quantification of resonant refractive index sensors," Opt. exp. 16, 1020-1028 (2008).

[33] G.N. Tsigaridas, "A study on refractive index sensors based on optical micro-ring resonators," Photonic Sens, 7: 217 2017.

[34] C. Yim, M. O'Brien N. McEvoy, S. Winters, I. Mirza, J. G. Lunney, and G. S. Duesberg, "Investigation of the optical properties of MoS2 thin films using spectroscopic ellipsometry" Appl. Phys. Lett. 104, 103114 (2014).

[35] D. T. Pierce and W. E. Spicer." Electronic structure of amorphous Si from photoemission and optical studies," Phys. Rev. B 5, 3017-3029 (1972).